# Examining EAP Students' AI Disclosure Intention: A Cognition–Affect–Conation Perspective

**First Author and Corresponding Author**
Yiran Du
University of Cambridge
yd392@cam.ac.uk

**Second Author**
Huimin He
Huimin.he@xjtlu.edu.cn
Xi'an Jiaotong-Liverpool University, Suzhou, China

**Abstract**
The growing use of generative artificial intelligence (AI) in academic writing has raised increasing concerns regarding transparency and academic integrity in higher education. This study examines the psychological factors influencing English for Academic Purposes (EAP) students' intention to disclose their use of AI tools. Drawing on the cognition–affect–conation framework, the study proposes a model integrating both enabling and inhibiting factors shaping disclosure intention. A sequential explanatory mixed-methods design was employed. Quantitative data from 324 EAP students at an English-medium instruction university in China were analysed using structural equation modelling, followed by semi-structured interviews with 15 students to further interpret the findings. The quantitative results indicate that psychological safety positively predicts AI disclosure intention, whereas fear of negative evaluation negatively predicts it. The qualitative findings further reveal that supportive teacher practices and clear guidance foster psychological safety, while policy ambiguity and reputational concerns intensify fear of negative evaluation and discourage disclosure. These findings highlight the importance of clear institutional policies and supportive pedagogical environments in promoting transparent AI use.

## 1. Introduction

The rapid development of artificial intelligence (AI), particularly generative AI technologies, has significantly transformed language learning and academic writing practices in higher education. In English for Academic Purposes (EAP) contexts, AI-powered tools such as natural language processing systems and generative writing assistants can support learners in tasks including idea generation, paraphrasing, grammar correction, and summarisation (Y. Du et al., 2025). These technologies may be particularly valuable for EAP students who are simultaneously developing academic literacy and second-language proficiency, as AI tools can provide linguistic scaffolding that supports engagement with complex academic writing conventions (Zou et al., 2024). Consequently, AI technologies are increasingly integrated into language learning environments, creating new opportunities for supporting academic writing development.

Despite these potential benefits, the growing use of AI in academic writing has raised important pedagogical and ethical concerns. Educators and institutions have expressed increasing concern regarding students' overreliance on AI-generated content and the implications for academic integrity and independent learning (Aljohani, 2026; Gruenhagen et al., 2024). In response, many universities have begun introducing policies that require students to disclose their use of AI tools in academic work to ensure transparency and responsible use (Garcia Ramos, 2025). AI disclosure, defined as students' willingness to report or acknowledge their use of AI tools in completing academic tasks, has therefore become an emerging issue within higher education (Ans et al., 2026; Zhang et al., 2025). Understanding the factors that influence students' willingness to disclose AI use is increasingly important as institutions attempt to balance technological innovation with academic integrity.

However, empirical research on AI disclosure intention remains limited, particularly in EAP contexts where students may encounter unfamiliar academic norms and expectations regarding AI-assisted writing. Prior studies have primarily focused on students' acceptance or use of AI technologies, with less attention given to the psychological mechanisms underlying disclosure behaviour (Li et al., 2025; Zhou & Wu, 2025). Drawing on the cognition–affect–conation framework, which conceptualises behaviour as a process shaped by cognitive evaluations and affective responses (Zhou & Zhang, 2024), the present study investigates the factors influencing EAP students' AI disclosure intention. By examining both enabling and inhibiting psychological factors, this research aims to provide a more comprehensive understanding of how students navigate emerging norms of transparency and responsible AI use in academic writing.

## 2. Literature Review
### 2.1 AI in EAP Education
Artificial intelligence (AI) technologies have become increasingly visible in English for Academic Purposes (EAP) education, particularly with the development of natural language processing systems and generative AI tools capable of producing extended academic discourse (Wang et al., 2026). These technologies can assist learners in a range of language-related tasks, including idea generation, vocabulary expansion, paraphrasing, summarisation, and grammatical correction (C. Du et al., 2025). For EAP students who are simultaneously developing academic literacy and second-language proficiency, AI tools can function as supplementary linguistic support that facilitates engagement with academic writing conventions (Wang et al., 2024).

From a pedagogical perspective, AI-based writing tools may support the writing process by providing immediate feedback and linguistic scaffolding (Y. Du et al., 2026). Research in second-language writing suggests that technological tools can help learners manage lower-level language concerns, thereby allowing greater attention to higher-order aspects of writing such as argumentation, organisation, and rhetorical structure (Lee et al., 2026). In EAP contexts, where students are often required to produce discipline-specific texts in English, AI systems may therefore serve as an auxiliary resource that assists with drafting, revision, and comprehension of academic texts (He & Du, 2024).

However, the growing presence of generative AI in academic writing also raises pedagogical and ethical challenges. Educators have expressed concerns regarding overreliance on AI-generated content, potential erosion of students' independent writing development, and ambiguity surrounding acceptable versus inappropriate AI use (Aljohani, 2026). As higher education institutions continue to formulate policies governing AI-assisted writing, issues of transparency and responsible use have become increasingly salient (Weng & Fu, 2025). Consequently, understanding how EAP students interact with AI tools in their learning processes has become an important area of inquiry within EAP and educational technology research.

### 2.2 AI Disclosure Intention
AI disclosure intention refers to students' willingness to acknowledge or report their use of artificial intelligence tools when completing academic tasks (Ans et al., 2026). With the rapid adoption of generative AI in higher education, many institutions have begun to require transparency regarding the use of such tools in academic work (Garcia Ramos, 2025). Disclosure is therefore increasingly framed as a component of academic integrity, allowing instructors to distinguish between legitimate technological assistance and inappropriate reliance on AI-generated content (Peng & Wan, 2025).

Scholars commonly conceptualise disclosure intention within broader frameworks of behavioural intention, which emphasise the psychological processes underlying decision-making (Zhou & Wu, 2025a). In this perspective, intention represents the motivational state that precedes actual behaviour. Students' decisions to disclose AI use may therefore be influenced by multiple psychological factors, including their beliefs about the usefulness and legitimacy of AI tools, their emotional responses such as trust or anxiety, and their perceived obligations to comply with institutional norms (Zhang et al., 2025). These cognitive and affective evaluations could shape students' willingness to report their AI usage in academic work.

Despite increasing institutional attention to AI transparency, empirical research on students' AI disclosure intention remains limited. Existing studies have primarily focused on attitudes toward AI use or concerns about academic integrity, while fewer investigations have examined the psychological mechanisms that influence disclosure behaviour (B. Li et al., 2025). This gap is particularly relevant in EAP contexts, where international students may encounter unfamiliar academic conventions and expectations regarding AI use. Examining AI disclosure intention among EAP students therefore contributes to a deeper understanding of how learners navigate emerging norms of responsible AI use in academic writing.

## 3. Theoretical Framework

This study adopts a cognition–affect–conation framework to examine EAP students' AI disclosure intention. The framework conceptualises behaviour as a sequential process in which cognitive evaluations influence affective responses, which subsequently shape behavioural intentions (Zhou & Zhang, 2024). It has been widely applied in behavioural and educational research to explain how individuals' beliefs and emotions jointly shape decision-making processes (Zhou & Zhang, 2025). In the context of AI-assisted academic writing, students' decisions about whether to disclose their use of AI tools are likely influenced not only by their perceptions of institutional conditions but also by their emotional reactions to potential consequences.

To capture these dynamics, the present study incorporates both enabling and inhibiting factors across the cognitive and affective stages. Considering both types of factors is important because students' disclosure decisions may be simultaneously encouraged by supportive conditions and constrained by perceived risks or concerns. Including both enablers and inhibitors therefore allows a more comprehensive understanding of the psychological mechanisms underlying disclosure intention, as students' behaviour may be shaped by the interaction between positive and negative perceptions and emotions.

Within the proposed conceptual model (see Figure 1), the cognitive enablers include perceived fairness, perceived teacher support, and self-efficacy, whereas the cognitive inhibitors include perceived uncertainty, perceived susceptibility, and perceived risk. At the affective level, psychological safety functions as an enabling factor that may facilitate disclosure intention, while fear of negative evaluation acts as an inhibiting factor that may discourage disclosure. These affective responses subsequently influence the conative outcome, namely AI disclosure intention.

**Figure 1. The Conceptual Model**

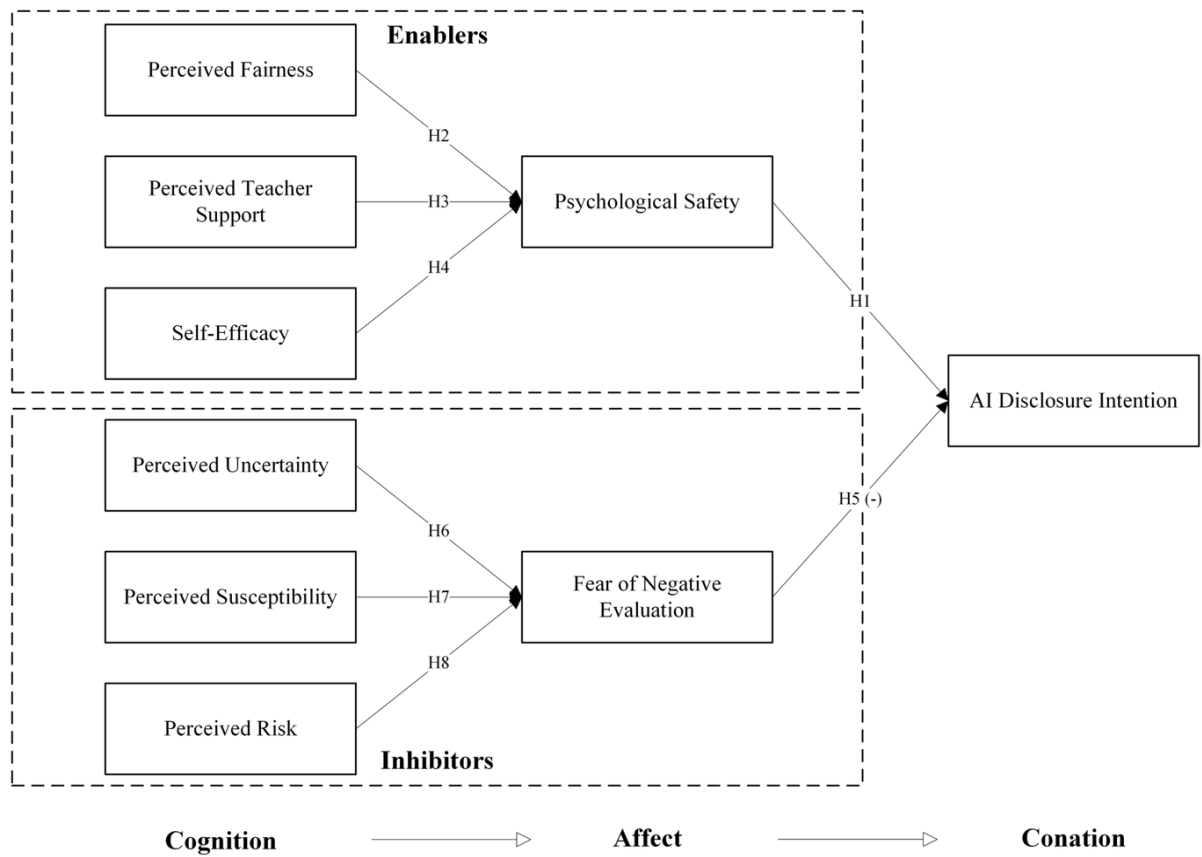

## 4. Research Questions and Hypotheses
### 4.1 Enablers of AI Disclosure Intention
Based on the conceptual model, the first research question (RQ1) asks: *How do psychological safety, perceived fairness, perceived teacher support, and self-efficacy affect EAP students' AI disclosure intention?* Psychological safety refers to the perception that individuals can acknowledge their behaviours without fear of negative consequences (Edmondson & Bransby, 2023). The concept originates from organisational and educational research, which suggests that psychologically safe environments encourage openness, transparency, and responsible behaviour (Vella et al., 2024). In learning contexts, students who feel safe are more willing to express ideas, admit uncertainties, and report their learning practices (Somerville et al., 2023). Empirical studies in educational settings have shown that psychological safety facilitates honest communication and ethical behaviour (McGuire, 2025). Therefore, when students perceive that disclosing their use of AI tools will not lead to unfair judgement or punishment, they may feel more comfortable acknowledging such use in their academic work. Accordingly, psychological safety is expected to positively influence EAP students' AI disclosure intention.

Perceived fairness may serve as an important cognitive antecedent of psychological safety. Perceived fairness refers to students' evaluations of whether institutional policies and classroom practices are reasonable, transparent, and consistently implemented (Lünich et al., 2024). According to theories of organisational justice, fairness perceptions shape individuals' trust in institutions and their willingness to comply with rules. When rules are perceived as fair, individuals are more likely to view them as legitimate and feel secure in following them (Malhotra et al., 2022). Empirical research in educational contexts similarly indicates that fair evaluation practices and transparent policies enhance students' trust in instructors and institutions (Chambers et al., 2023). In the context of AI-assisted learning, when students perceive AI-related policies as fair and consistently applied, they may experience greater psychological safety when reporting their AI use.

Perceived teacher support is another cognitive factor that may enhance psychological safety. Teachers play a central role in establishing classroom norms and interpreting institutional policies (Tao et al., 2022). Educational research has consistently shown that supportive teacher–student relationships promote students' engagement, trust, and willingness to communicate openly (Held & Mori, 2024). When instructors provide clear guidance about acceptable AI use and demonstrate openness toward AI-assisted learning, students may interpret the learning environment as supportive rather than punitive. Such perceptions may reduce students' concerns about negative evaluation and strengthen their sense of safety when acknowledging their use of AI tools.

Finally, self-efficacy may also contribute to students' psychological safety in the disclosure process. Self-efficacy refers to individuals' belief in their capability to successfully perform academic tasks (Waddington, 2023). According to social cognitive theory, self-efficacy influences how individuals approach challenges and respond to potential risks (Graham, 2022). Students with higher academic self-efficacy tend to display greater confidence, persistence, and resilience in learning situations. Empirical research has also shown that students with stronger self-efficacy are less likely to experience anxiety related to evaluation (Bergdahl & Sjöberg, 2025). In the context of AI-assisted writing, students who are confident in their academic abilities may feel less concerned that acknowledging AI assistance will undermine perceptions of their competence, thereby increasing their psychological comfort in disclosing AI use (Ulfert-Blank & Schmidt, 2022). Accordingly, the following hypotheses are proposed:

H1: Psychological safety positively predicts AI disclosure intention.
H2: Perceived fairness positively predicts psychological safety.
H3: Perceived teacher support positively predicts psychological safety.
H4: Self-efficacy positively predicts psychological safety.

### 4.2 Inhibitors of AI Disclosure Intention

Based on the conceptual model, the second research question (RQ2) asks: *How do fear of negative evaluation, perceived uncertainty, perceived susceptibility, and perceived risk affect EAP students' AI disclosure intention?* Fear of negative evaluation refers to individuals' apprehension about being judged unfavourably by others (Piko et al., 2024). The concept has been widely examined in social and educational psychology, where it is associated with anxiety about criticism and concern over others' perceptions (Liu et al., 2022). In academic settings, students who fear negative evaluation may avoid behaviours that could potentially damage their academic image (Jia & Yue, 2023). When students believe that acknowledging their use of AI tools might lead to criticism from teachers or peers, they may become reluctant to disclose such use. Therefore, fear of negative evaluation is expected to reduce EAP students' AI disclosure intention.

Perceived uncertainty may function as a cognitive factor that increases fear of negative evaluation. Uncertainty refers to students' perceptions that institutional policies or classroom expectations regarding AI use are unclear or ambiguous. Research on emerging technologies in education suggests that unclear rules often create confusion about acceptable practices (Usher & Barak, 2024). When students are uncertain about whether certain forms of AI use are permitted, they may worry that their actions could be interpreted negatively by instructors (Zhu et al., 2025). This ambiguity may increase students' concerns about potential judgement and thereby strengthen their fear of negative evaluation (Stone, 2025).

Another cognitive inhibitor is perceived susceptibility, which refers to students' perceptions of the likelihood that their AI use may be detected or questioned by instructors. Drawing from theories of risk perception and protection motivation, individuals who believe they are more likely to be monitored or exposed tend to experience stronger concerns about potential consequences (Fischer-Preßler et al., 2022). In academic contexts, when students believe that instructors or institutions can identify AI-assisted writing, they may anticipate criticism or disciplinary responses (Kim et al., 2025). Such perceptions may heighten their sensitivity to negative judgement and contribute to fear of negative evaluation.

Finally, perceived risk may also increase fear of negative evaluation. Perceived risk refers to students' assessment of the potential negative outcomes associated with disclosing AI use, such as lower grades, academic penalties, or damage to their academic reputation (Wu et al., 2022). Research on academic integrity indicates that students' behavioural decisions are strongly influenced by their perceptions of potential costs and consequences (Gruenhagen et al., 2024). When students perceive higher risks associated with AI use disclosure, they may experience stronger concerns about negative judgement, which may further intensify their fear of negative evaluation (Balalle & Pannilage, 2025). Accordingly, the following hypotheses are proposed:

H5: Fear of negative evaluation negatively predicts AI disclosure intention.
H6: Perceived uncertainty positively predicts fear of negative evaluation.
H7: Perceived susceptibility positively predicts fear of negative evaluation.
H8: Perceived risk positively predicts fear of negative evaluation.

## 5. Methods
### 5.1 Research Design
This study adopted a sequential explanatory mixed-methods design, in which quantitative data collection and analysis were followed by qualitative inquiry to further interpret the quantitative findings (Cohen et al., 2018). In the first phase, a questionnaire survey was administered to examine the relationships among the constructs in the proposed conceptual model of AI disclosure intention. The quantitative results provided an overview of students' perceptions and the hypothesised relationships among cognitive, affective, and behavioural factors. In the second phase, semi-structured interviews were conducted with a purposively selected subset of participants to gain deeper insights into students' experiences and perspectives regarding AI use and disclosure in academic writing. The qualitative findings were used to elaborate and contextualise the quantitative results, thereby providing a more comprehensive understanding of the factors influencing EAP students' AI disclosure intention.

### 5.2 Participants
A total of 348 students initially responded to the questionnaire. After data screening, 24 responses were excluded due to careless responding (e.g., patterned responses or failure to pass attention checks), resulting in a final valid sample of 324 participants for analysis. Participants were recruited from an English-medium instruction (EMI) university in China through announcements made in compulsory English for Academic Purposes (EAP) classes. At this university, EAP courses are required during the first two years of undergraduate study. The survey invitation was distributed during class time, and students who were willing to participate completed the online questionnaire voluntarily. The demographic characteristics of the questionnaire participants are summarised in Appendix A. The sample consisted of 182 female students (56.2%) and 142 male students (43.8%). In terms of age distribution, 40.7% of participants were aged 18–20, 48.8% were aged 21–23, and 10.5% were aged 24 or above. Regarding disciplinary background, 60.5% of respondents were enrolled in non-STEM programmes, whereas 39.5% were studying in STEM fields.

For the qualitative phase, interview participants were selected from questionnaire respondents who indicated their willingness to participate in a follow-up interview. A purposeful sampling strategy was employed, prioritising variation in AI disclosure intention and subsequently seeking diversity in gender, age, and disciplinary background (STEM vs non-STEM). Disclosure intention levels were determined based on participants' scores on the AI disclosure intention scale in the questionnaire. For each respondent, a mean score across the disclosure intention items was calculated. Participants whose scores fell within the top 25% of the distribution were classified as having high disclosure intention, whereas those within the bottom 25% were classified as having low disclosure intention. Students were first selected to ensure representation from both high and low disclosure intention groups, after which variation in gender, age, and field of study was considered. In total, 15 students participated in the semi-structured interviews. The demographic characteristics of the interview participants are presented in Appendix B. Interviews were conducted until thematic saturation was reached, meaning that no substantially new themes emerged from additional interviews.

This study followed established ethical guidelines for research involving human participants. Participation in both the questionnaire and interview phases was voluntary, and informed consent was obtained from all participants prior to data collection. Participants were informed about the purpose of the study, their right to withdraw at any time, and the confidentiality of their responses. All questionnaire responses were collected anonymously, and interview data were anonymised using participant codes to protect participants' identities.

### 5.3 Questionnaires

Data were collected using a structured questionnaire designed to measure the constructs in the proposed conceptual model. The questionnaire included nine constructs: perceived fairness (Chambers et al., 2023), perceived teacher support (Chiu et al., 2023), self-efficacy (Bai & Wang, 2023), perceived uncertainty (Usman et al., 2021), perceived susceptibility (Siani et al., 2024), perceived risk (W. Li, 2025), psychological safety (McGuire, 2025), fear of negative evaluation (Jia & Yue, 2023), and AI disclosure intention (Zhou & Wu, 2025b). All items were adapted from established scales in prior research and modified to fit the context of AI use in academic writing (Y. Du, 2024). The questionnaire was administered in Chinese to ensure participants' comprehension, and a translation and back-translation procedure was conducted to maintain equivalence between the English and Chinese versions (Brislin, 1970). Prior to the main data collection, a pilot study with 30 students was conducted to assess the clarity and appropriateness of the questionnaire items, and minor wording revisions were made based on their feedback. The questionnaire also collected participants' demographic information, including gender, age, and field of study, and invited respondents to indicate their willingness to participate in a follow-up interview. All items were measured using a five-point Likert scale ranging from 1 (strongly disagree) to 5 (strongly agree). The full list of constructs and measurement items is presented in Appendix C.

### 5.4 Semi-Structured Interview

To complement the questionnaire findings, semi-structured interviews were conducted to obtain a deeper understanding of students' perceptions and experiences regarding AI use and disclosure in academic writing. The interview protocol was developed based on the hypotheses of the proposed conceptual model, with questions designed to explore participants' views on factors such as perceived fairness, teacher support, self-efficacy, perceived uncertainty, perceived susceptibility, perceived risk, psychological safety, and fear of negative evaluation in relation to their willingness to disclose AI use. Semi-structured interviews were chosen to allow flexibility for participants to elaborate on their experiences while ensuring that key topics relevant to the research questions were consistently addressed. Participants who had indicated their willingness to participate in the questionnaire were contacted and invited to take part in the interviews. The interviews were conducted individually in Chinese and lasted approximately 30–45 minutes. After obtaining informed consent, interviews followed the interview protocol, with additional probing questions used when necessary to clarify responses and encourage elaboration. The full interview protocol is presented in Appendix D.

### 5.5 Data Analysis

The quantitative data were analysed using R following a two-step structural equation modelling (SEM) procedure (Whittaker & Schumacker, 2022). First, descriptive statistics and preliminary data screening were conducted to examine missing values, outliers, and the distribution of the variables. Skewness and kurtosis values were inspected to assess normality. Subsequently, confirmatory factor analysis (CFA) was performed to evaluate the measurement model, including model fit, indicator reliability, internal consistency reliability, convergent validity, and discriminant validity. Reliability was assessed using Cronbach's α and composite reliability (CR), while convergent validity was evaluated through standardised factor loadings and average variance extracted (AVE). Discriminant validity was examined using the Fornell–Larcker criterion. After establishing satisfactory measurement properties, the structural model was tested to examine the hypothesised relationships among the constructs within the cognition–affect–conation framework. Path coefficients and significance levels were used to determine whether the proposed hypotheses were supported.

The qualitative interview data were analysed using thematic analysis following a deductive–inductive approach (King et al., 2019). Initially, a deductive coding framework was developed based on the constructs in the proposed conceptual model. The interview transcripts were then coded using this preliminary framework while allowing new themes to emerge inductively from the data. Two researchers independently coded a subset of the transcripts to establish coding reliability. Inter-coder agreement was assessed using Cohen's kappa, which indicated substantial agreement ($\kappa = 0.83$) (McHugh, 2012). After resolving discrepancies through discussion, the coding scheme was refined and applied to the remaining transcripts. The resulting codes were then grouped into broader themes that captured students' cognitive evaluations, affective responses, and behavioural intentions related to AI disclosure.

## 6. Results
### 5.1 Quantitative Results
Appendix E presents the descriptive statistics of the study constructs, including the means and standard deviations for each variable. To assess the normality of the data, skewness and kurtosis values were examined. Following commonly accepted guidelines, absolute skewness values below 2 and kurtosis values below 7 indicate acceptable univariate normality. As shown in Appendix E, all constructs fell within these thresholds, suggesting that the assumption of normality was satisfied and that the data were appropriate for subsequent statistical analyses.

The measurement model was evaluated in terms of model fit, reliability, convergent validity, and discriminant validity. As shown in Appendix F, the measurement model demonstrated an acceptable fit to the data, with all fit indices meeting the recommended thresholds. As presented in Appendix G, all standardised factor loadings exceeded the recommended threshold of 0.70, indicating satisfactory indicator reliability. Cronbach's α values ranged from 0.82 to 0.89, and composite reliability (CR) values ranged from 0.85 to 0.91, both exceeding the recommended threshold of 0.70, which demonstrates adequate internal consistency reliability. In addition, the average variance extracted (AVE) values ranged from 0.65 to 0.77, surpassing the recommended threshold of 0.50 and indicating satisfactory convergent validity. Discriminant validity was assessed using the Fornell–Larcker criterion (Appendix H). The square roots of the AVE values for each construct were greater than their correlations with other constructs, suggesting that adequate discriminant validity was established. Overall, these results indicate that the measurement model exhibited satisfactory reliability and validity.

The structural model demonstrated an acceptable fit to the data, with all model fit indices meeting the recommended thresholds (Appendix F). The structural path estimates are presented in Appendix I and Figure 2. Psychological safety had a significant positive effect on AI disclosure intention ($\beta = 0.42$, $p < .001$), supporting H1. Perceived fairness ($\beta = 0.28$, $p < .001$), perceived teacher support ($\beta = 0.31$, $p < .001$), and self-efficacy ($\beta = 0.24$, $p < .01$) were found to positively predict psychological safety, supporting H2, H3, and H4. In addition, fear of negative evaluation had a significant negative effect on AI disclosure intention ($\beta = -0.29$, $p < .001$), supporting H5. Regarding the antecedents of fear of negative evaluation, perceived uncertainty ($\beta = 0.33$, $p < .001$), perceived susceptibility ($\beta = 0.18$, $p < .05$), and perceived risk ($\beta = 0.37$, $p < .001$) all showed significant positive effects, supporting H6, H7, and H8.

**Figure 2. Structural Model Results**

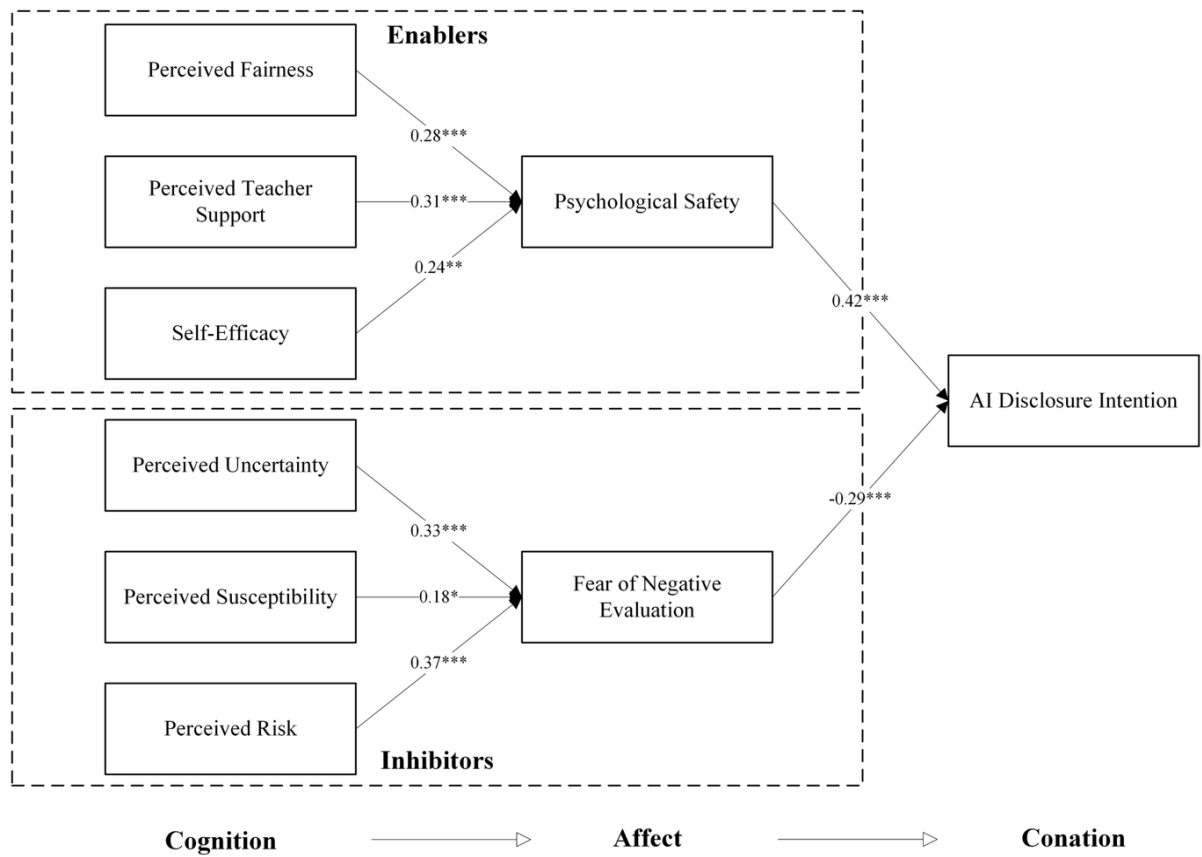

Note. Statistical significance is denoted as *** *p* < .001, ** *p* < .01, * *p* < .05.

## 5.2 Qualitative Results

The qualitative interview data were analysed thematically to provide deeper insight into the psychological mechanisms underlying students' AI disclosure intention. The analysis generated three interrelated themes: (1) supportive learning environments and psychological safety, (2) policy ambiguity and institutional uncertainty, and (3) reputational concerns and fear of negative evaluation. These themes broadly align with the cognition–affect–conation framework used in this study, illustrating how students' cognitive evaluations of institutional conditions shape their emotional responses and, ultimately, their behavioural intentions regarding AI disclosure.

The first theme highlights the role of supportive learning environments in fostering psychological safety and encouraging disclosure. Many participants with relatively high disclosure intention described teacher attitudes as a critical factor shaping their willingness to acknowledge AI use. When instructors explicitly discussed acceptable forms of AI assistance and framed AI as a legitimate learning tool, students reported feeling more comfortable being transparent about their practices. As one student explained, *"If the teacher clearly says AI can help with ideas or grammar, then I think it is okay to mention it in the assignment"* (P2). Similarly, another participant noted that openness from instructors reduced anxiety about potential consequences: *"When teachers treat AI as part of learning instead of cheating, I feel safer saying that I used it"* (P5). These accounts suggest that supportive teacher–student relationships can cultivate psychological safety by signalling that disclosure will not necessarily lead to negative judgement. However, several students also indicated that such support varied considerably across courses, suggesting that disclosure norms are often shaped more by individual instructors than by consistent institutional policies.

The second theme concerns policy ambiguity and uncertainty surrounding acceptable AI use, which many students perceived as a barrier to disclosure. Participants frequently reported confusion about the boundaries between legitimate AI assistance and academic misconduct. For example, one participant remarked, *"We know AI should not write the whole essay, but the rules do not really say how much help*

*is acceptable"* (P9). This ambiguity appeared to create a sense of cognitive uncertainty that complicated students' decision-making. Some students worried that disclosure might invite additional scrutiny from instructors, even when AI use was limited. As another participant stated, *"If I write that I used AI, maybe the teacher will check my work more carefully or question it"* (P10). Such perceptions suggest that unclear policies may inadvertently discourage transparency, as students may perceive non-disclosure as a safer option in contexts where expectations remain uncertain.

The third theme relates to reputational concerns and fear of negative evaluation, which were particularly evident among participants with lower disclosure intention. Several students expressed concern that admitting AI use could damage their academic image or lead teachers to question their competence. One participant explained, *"If I say I used AI, the teacher might think I cannot write well by myself"* (P12). Another student similarly commented that AI use might be interpreted negatively regardless of institutional policies: *"Even if AI is allowed, teachers might still think you are lazy or relying too much on technology"* (P13). These responses illustrate how disclosure decisions are influenced not only by formal rules but also by perceived social norms within academic environments. The persistence of stigma surrounding AI-assisted writing appears to intensify students' fear of negative evaluation, thereby reducing their willingness to disclose AI use.

Taken together, these qualitative findings provide contextual depth to the quantitative results reported earlier. While the survey analysis indicated that psychological safety positively predicts AI disclosure intention and fear of negative evaluation negatively predicts it, the interview data reveal the situational conditions through which these affective states emerge. Supportive teacher practices and clear guidance can enhance students' sense of safety, whereas policy ambiguity and reputational concerns may heighten anxiety about potential judgement. Importantly, the findings also suggest that these influences often coexist within the same institutional environment. Students may simultaneously recognise the pedagogical value of AI while remaining uncertain about how their disclosure will be interpreted. This tension highlights the need for clearer institutional guidance and more consistent pedagogical practices if universities aim to encourage transparent and responsible AI use in academic writing.

## 7. Discussion
### 7.1 Enablers of AI Disclosure Intention

The findings indicate that psychological safety plays a central role in enabling students' intention to disclose AI use in academic writing. The quantitative results show that when students perceive the learning environment as psychologically safe, they are more willing to report their AI use. The interview data provide further insight into this relationship: several students explained that they felt more comfortable acknowledging AI assistance when instructors framed AI as a learning tool rather than as misconduct. This pattern is consistent with prior research suggesting that psychologically safe environments promote openness and responsible behaviour in educational settings (Edmondson & Bransby, 2023; McGuire, 2025). At the same time, the qualitative accounts indicate that psychological safety is fragile and highly dependent on classroom practices, suggesting that institutional encouragement of disclosure may not automatically translate into students' perceived safety.

The results also highlight the importance of perceived fairness and teacher support as cognitive conditions that foster psychological safety. When students perceive AI-related policies as fair and consistently implemented, they are more likely to interpret disclosure requirements as legitimate. This finding aligns with organisational justice literature indicating that fairness perceptions strengthen trust in institutional rules and encourage compliance (Malhotra et al., 2022; Lünich et al., 2024). The interview data further suggest that fairness is closely tied to clarity: students reported greater willingness to disclose AI use when instructors clearly explained acceptable practices. However, the findings also reveal that disclosure norms are often shaped by individual teachers rather than consistent institutional policies, which may create uneven disclosure expectations across courses.

Teacher support further contributes to disclosure intention by shaping students' interpretation of the learning environment. Consistent with previous research showing that supportive teacher–student relationships promote engagement and open communication (Held & Mori, 2024; Tao et al., 2022), students in this study reported that teachers' attitudes toward AI strongly influenced their willingness to disclose its use. When instructors openly discussed AI as a legitimate academic tool, students perceived disclosure as less risky. Nevertheless, the qualitative findings suggest that supportive practices vary across classrooms, indicating that teacher-level differences may significantly shape students' disclosure behaviour.

Finally, self-efficacy appears to facilitate disclosure by reducing concerns about reputational judgement. Students who feel confident in their academic abilities may be less worried that acknowledging AI assistance will undermine perceptions of their competence. This interpretation is consistent with social cognitive perspectives that link self-efficacy with reduced evaluation anxiety (Graham, 2022; Waddington, 2023). However, the qualitative evidence suggests that students with lower confidence may remain reluctant to disclose AI use, even in relatively supportive environments. This finding implies that disclosure policies alone may not be sufficient; fostering student confidence and clarifying the role of AI in learning may be equally important for encouraging transparent AI use.

### 7.2 Inhibitors of AI Disclosure Intention
The findings indicate that fear of negative evaluation functions as a key affective barrier to students' AI disclosure intention. The quantitative results show that students who experience stronger concerns about being judged unfavourably are less willing to report their AI use. This pattern is consistent with research in educational psychology suggesting that fear of negative evaluation can discourage behaviours that might threaten one's academic image (Liu et al., 2022; Jia & Yue, 2023). The interview data provide further context for this relationship. Several participants expressed concern that admitting AI use might lead teachers to question their competence or effort, even when AI was used only for limited support. These perceptions suggest that disclosure decisions are shaped not only by formal institutional policies but also by students' expectations of how their behaviour will be socially interpreted.

The results further show that perceived uncertainty surrounding AI policies contributes to this fear of negative evaluation. When students perceive institutional guidelines as ambiguous, they may become unsure about whether their AI use will be considered acceptable. Such uncertainty can increase anxiety about potential judgement from instructors. Previous research on emerging technologies in education similarly indicates that unclear rules often create confusion about acceptable practices and discourage transparent behaviour (Stone, 2025; Usher & Barak, 2024). The interview findings reinforce this point: several students reported that the lack of explicit boundaries regarding acceptable AI assistance made them hesitant to disclose its use.

Perceived susceptibility also appears to influence fear of negative evaluation. The quantitative results indicate that when students believe their AI use is likely to be detected or questioned, their concern about negative judgement increases. This observation aligns with protection motivation perspectives, which suggest that perceived monitoring or exposure can heighten anxiety about potential consequences (Fischer-Preßler et al., 2022). Interview responses illustrate this dynamic: some students noted that disclosing AI use might prompt instructors to scrutinise their work more closely, thereby increasing the likelihood of criticism.

Finally, perceived risk plays an important role in shaping students' reluctance to disclose AI use. When students believe that disclosure could lead to negative outcomes, such as lower grades or reputational damage, they may prefer to avoid reporting AI assistance altogether. Prior research on academic integrity similarly indicates that perceived costs and potential sanctions strongly influence students' behavioural decisions (Gruenhagen et al., 2024; Wu et al., 2022). However, the qualitative findings suggest that these risks are often interpreted subjectively rather than based on explicit institutional rules. As a result, even in environments where AI use may be partially permitted, students may still perceive disclosure as a potentially risky action.

### 7.3 Theoretical and Practical Implications
This study contributes to the emerging literature on AI use in higher education by extending the cognition–affect–conation framework to the context of AI disclosure behaviour. While prior research has largely focused on students' acceptance or use of AI tools, the present findings highlight disclosure as a distinct behavioural outcome shaped by both cognitive evaluations and affective responses (Zhou & Zhang, 2024; Zhou & Wu, 2025). By integrating enabling and inhibiting factors within a single model, the study demonstrates that disclosure intention is influenced by the interaction between supportive perceptions, such as fairness and teacher support, and constraining perceptions, such as uncertainty and perceived risk. The mixed-methods design further strengthens this theoretical contribution by showing how quantitative relationships correspond to students' lived experiences, thereby illustrating the psychological processes through which institutional conditions and classroom environments translate into disclosure behaviour.

The findings also offer practical implications for universities seeking to promote transparent and responsible AI use in academic work. First, the results suggest that institutional policies alone are insufficient to encourage disclosure unless they are perceived as fair, clearly communicated, and consistently implemented. Providing explicit guidance on acceptable AI practices may reduce uncertainty and strengthen students' sense of psychological safety. Second, the findings highlight the influential role of instructors in shaping disclosure norms. When teachers adopt an open and pedagogically oriented approach to AI use, students appear more willing to acknowledge their use of AI tools. Consequently, institutions may need to support educators through training and guidance on how to discuss and manage AI use in the classroom. More broadly, fostering supportive learning environments that emphasise transparency rather than punishment may be essential for encouraging responsible AI engagement among students.

### 7.4 Limitations and Future Directions
One limitation of this study concerns its contextual scope and reliance on self-reported perceptions. The data were collected from students at a single English-medium instruction university, which may limit the transferability of the findings to other institutional or cultural contexts where AI policies, academic integrity norms, and pedagogical practices differ. In addition, the study measured disclosure intention rather than actual disclosure behaviour. Although intention is widely considered a proximal predictor of behaviour in behavioural research, self-reported responses may still be influenced by social desirability or students' interpretations of institutional expectations. As a result, the findings should be interpreted as reflecting students' perceived motivations and concerns rather than direct evidence of their disclosure practices.

Future research could address these limitations by expanding both the methodological and contextual scope of investigation. Comparative studies across universities or national contexts would help clarify how different policy environments and academic cultures shape students' AI disclosure behaviour. Longitudinal research may also be useful for examining how disclosure intentions evolve as institutional guidelines become clearer and as students gain more experience with AI tools. In addition, future studies could explore the role of instructional practices and AI literacy initiatives in shaping students' perceptions of psychological safety and transparency, thereby providing deeper insight into how educational institutions can encourage responsible and open engagement with AI in academic writing.

### 8. Conclusion
This study examined the psychological factors influencing EAP students' AI disclosure intention through a cognition–affect–conation framework using a sequential explanatory mixed-methods design. The findings indicate that disclosure intention is shaped by the interaction between enabling and inhibiting mechanisms. Psychological safety emerged as a key affective driver encouraging disclosure, while perceived fairness, teacher support, and self-efficacy functioned as cognitive conditions that foster this sense of safety. Conversely, fear of negative evaluation acted as a major barrier to disclosure, with perceived uncertainty, susceptibility, and risk intensifying students' concerns about potential judgement. The qualitative findings further revealed that these psychological processes are closely tied to classroom practices and policy clarity, highlighting how institutional ambiguity and reputational concerns may discourage transparency even when AI use is partially accepted. Overall, the study underscores the importance of supportive pedagogical environments and clear institutional guidance in promoting transparent and responsible AI use in EAP education.

**Declarations**
**Acknowledgements**
None.

**Competing Interests**
None.

**Funding**
None.

**Data Availability Statement**
Data are available from the corresponding author upon reasonable request.

**Appendices**

**Appendix A. Questionnaire Participant Characteristics (*N* = 324)**

| Characteristic | Category | n | % |
| --- | --- | --- | --- |
| Gender | Male | 142 | 43.8 |
|  | Female | 182 | 56.2 |
| Age | 18–20 | 132 | 40.7 |
|  | 21–23 | 158 | 48.8 |
|  | 24 and above | 34 | 10.5 |
| Field of Study | STEM | 128 | 39.5 |
|  | Non-STEM | 196 | 60.5 |

**Appendix B. Interview Participant Characteristics (*N* = 15)**

| Participant | Gender | Age | Field of Study | AI Disclosure Intention Level |
| --- | --- | --- | --- | --- |
| P1 | Female | 19 | Non-STEM | High |
| P2 | Male | 20 | STEM | High |
| P3 | Female | 21 | Non-STEM | High |
| P4 | Male | 22 | STEM | High |
| P5 | Female | 20 | Non-STEM | High |
| P6 | Male | 21 | Non-STEM | High |
| P7 | Female | 19 | STEM | High |
| P8 | Male | 22 | Non-STEM | Low |
| P9 | Female | 21 | Non-STEM | Low |
| P10 | Male | 20 | STEM | Low |
| P11 | Female | 23 | Non-STEM | Low |
| P12 | Male | 21 | STEM | Low |
| P13 | Female | 22 | Non-STEM | Low |
| P14 | Male | 20 | Non-STEM | Low |
| P15 | Female | 24 | STEM | Low |

**Appendix C. Constructs and Measurement Items**

| Construct | Item | Measurement Item (English) | Measurement Item (Chinese) |
| --- | --- | --- | --- |
| Perceived Fairness | PF1 | The university's policies regarding AI use in academic work are fair. | 学校关于在学术作业中使用人工智能的政策是公平的。 |
| (Chambers et al., 2023) | PF2 | The rules about AI use in coursework are applied consistently by instructors. | 教师在课程作业中对人工智能使用规则的执行是一致的。 |
| | PF3 | The guidelines about AI use are reasonable. | 关于人工智能使用的指导原则是合理的。 |
| Perceived Teacher Support (Chiu et al., 2023) | PTS1 | My instructors provide clear guidance about acceptable AI use in coursework. | 我的教师会对课程作业中可接受的人工智能使用提供清晰指导。 |
| | PTS2 | My instructors are supportive when students discuss their use of AI tools. | 当学生讨论使用人工智能工具时，我的教师是支持的。 |
| | PTS3 | I feel comfortable asking my instructors questions about AI use in assignments. | 在作业中有关人工智能使用的问题上，我感到可以自在地向教师提问。 |
| Self-Efficacy (Bai & Wang, 2023) | SE1 | I am confident in my ability to complete academic writing tasks. | 我对自己完成学术写作任务的能力充满信心。 |
| | SE2 | I believe I can perform well in academic writing tasks. | 我相信自己能够在学术写作任务中表现良好。 |
| | SE3 | I feel capable of meeting the writing requirements of my courses. | 我认为自己能够达到课程的写作要求。 |
| Perceived Uncertainty (Usman et al., 2021) | PU1 | I am unsure about what kinds of AI use are allowed in academic assignments. | 我不确定在学术作业中哪些类型的人工智能使用是被允许的。 |
| | PU2 | The rules about AI use in coursework are sometimes unclear to me. | 对我来说，课程作业中关于人工智能使用的规则有时并不清晰。 |
| | PU3 | I find it difficult to determine whether certain AI-assisted practices are acceptable. | 我很难判断某些人工智能辅助的做法是否被允许。 |
| Perceived Susceptibility (Siani et al., 2024) | PS1 | Instructors are able to detect when AI tools have been used in academic writing. | 教师能够识别学术写作中是否使用了人工智能工具。 |
| | PS2 | It is likely that AI-assisted writing can be identified by instructors. | 人工智能辅助的写作很可能会被教师识别出来。 |
| | PS3 | Students who use AI tools without disclosure may be discovered. | 未披露使用人工智能工具的学生可能会被发现。 |
| Perceived Risk (Li, 2025) | PR1 | Disclosing my use of AI tools may negatively affect my grades. | 披露我使用人工智能工具可能会对我的成绩产生负面影响。 |
| | PR2 | Reporting AI use might lead instructors to evaluate my work more critically. | 报告使用人工智能可能会使教师更严格地评价我的作业。 |
| | PR3 | Disclosing AI use could have negative academic consequences. | 披露使用人工智能可能会带来不利的学术后果。 |
| Psychological Safety (McGuire, 2025) | PSY1 | I feel safe acknowledging my use of AI tools in academic work. | 在学术作业中承认使用人工智能工具时，我感到是安全的。 |
| | PSY2 | I believe I can disclose AI use without fear of negative judgement. | 我认为在披露人工智能使用时不必担心负面评价。 |

|  | PSY3 | I feel comfortable being transparent about AI use in my coursework. | 在课程作业中对人工智能使用保持透明让我感到自在。 |
| --- | --- | --- | --- |
| Fear of Negative Evaluation (Jia & Yue, 2023) | FNE1 | I worry that instructors will judge me negatively if I disclose AI use. | 我担心如果披露使用人工智能，教师会对我产生负面评价。 |
|  | FNE2 | I am concerned that admitting AI use may harm my academic image. | 我担心承认使用人工智能可能会损害我的学术形象。 |
|  | FNE3 | I feel anxious about how instructors might evaluate me if I report AI use. | 如果我报告使用人工智能，我会对教师如何评价我感到焦虑。 |
| AI Disclosure Intention (Zhou & Wu, 2025) | ADI1 | I intend to disclose my use of AI tools when completing academic assignments. | 在完成学术作业时，我打算披露自己使用了人工智能工具。 |
|  | ADI2 | I am willing to report AI assistance in my coursework when required. | 当有要求时，我愿意在课程作业中报告人工智能的辅助使用。 |
|  | ADI3 | I would acknowledge AI use if I used such tools while completing an assignment. | 如果我在完成作业时使用了人工智能工具，我会予以说明。 |

**Appendix D. Semi-Structured Interview Protocol**

| Hypothesis | Interview Question (English) | Interview Question (Chinese) |
| --- | --- | --- |
| H1: Psychological safety positively predicts AI disclosure intention. | How comfortable do you feel acknowledging or reporting your use of AI tools in academic assignments? Can you explain why? | 在学术作业中承认或报告使用人工智能工具时，你感到多大程度的自在或安全？可以解释一下原因吗？ |
| H2: Perceived fairness positively predicts psychological safety. | How fair do you think the university's policies or rules about AI use are? How does this influence whether you would disclose AI use? | 你认为学校关于人工智能使用的政策或规则是否公平？这是否会影响你是否披露使用人工智能？ |
| H3: Perceived teacher support positively predicts psychological safety. | How do your instructors respond to students' use of AI tools in coursework? Does their attitude affect whether you would disclose AI use? | 你的教师通常如何看待学生在课程作业中使用人工智能工具？他们的态度是否会影响你是否披露人工智能的使用？ |
| H4: Self-efficacy positively predicts psychological safety. | How confident do you feel about completing academic writing tasks without relying heavily on AI tools? Does this influence your willingness to disclose AI use? | 你对自己在不依赖人工智能工具的情况下完成学术写作任务有多大信心？这是否会影响你披露人工智能使用的意愿？ |
| H5: Fear of negative evaluation negatively predicts AI disclosure intention. | Do you worry about how instructors might judge you if you disclose that you used AI tools in an assignment? Why or why not? | 如果你在作业中披露使用了人工智能工具，你是否担心教师会如何评价你？为什么？ |
| H6: Perceived uncertainty positively predicts fear of negative evaluation. | How clear or unclear are the rules about AI use in your courses? How does this uncertainty affect your feelings about disclosing AI use? | 你认为课程中关于人工智能使用的规则是清晰还是不清晰？这种不确定性会如何影响你披露人工智能使用的想法？ |
| H7: Perceived susceptibility positively predicts fear of negative evaluation. | Do you think instructors can detect whether students use AI tools in their writing? How does this belief affect your behaviour? | 你认为教师能够识别学生在写作中是否使用了人工智能工具 |

| H8: Perceived risk positively predicts fear of negative evaluation. | What kinds of risks or consequences do you think might occur if you disclose your use of AI tools in academic work? | 吗？这种看法会如何影响你的行为？<br>如果你披露自己在学术作业中使用了人工智能工具，你认为可能会带来哪些风险或后果？ |
|---|---|---|

**Appendix E. Descriptive Statistics of the Constructs**

| Construct | *M* | SD | Skewness | Kurtosis |
|---|---|---|---|---|
| Perceived Fairness | 3.31 | 0.79 | -0.18 | -0.34 |
| Perceived Teacher Support | 3.26 | 0.82 | -0.11 | -0.41 |
| Self-Efficacy | 3.74 | 0.71 | -0.42 | 0.09 |
| Perceived Uncertainty | 3.58 | 0.83 | 0.21 | -0.29 |
| Perceived Susceptibility | 3.44 | 0.77 | 0.07 | -0.36 |
| Perceived Risk | 3.62 | 0.81 | 0.24 | -0.18 |
| Psychological Safety | 3.12 | 0.78 | -0.05 | -0.44 |
| Fear of Negative Evaluation | 3.67 | 0.80 | 0.31 | -0.21 |
| AI Disclosure Intention | 3.18 | 0.84 | 0.17 | -0.33 |

**Appendix F. Model Fit Indices**

| Fit Index | Threshold | Measurement Model | Structural Model |
|---|---|---|---|
| $\chi^2/df$ | < 3.00 | 2.11 | 2.27 |
| CFI | > 0.90 | 0.94 | 0.93 |
| TLI | > 0.90 | 0.93 | 0.92 |
| RMSEA | < 0.08 | 0.058 | 0.061 |
| SRMR | < 0.08 | 0.047 | 0.051 |

**Appendix G. Reliability and Convergent Validity**

| Construct | Item | Factor Loading | Cronbach's α | CR | AVE |
|---|---|---|---|---|---|
| Perceived Fairness | PF1 | 0.78 | 0.85 | 0.87 | 0.69 |
|  | PF2 | 0.86 |  |  |  |
|  | PF3 | 0.84 |  |  |  |
| Perceived Teacher Support | PTS1 | 0.80 | 0.86 | 0.88 | 0.71 |
|  | PTS2 | 0.87 |  |  |  |
|  | PTS3 | 0.83 |  |  |  |
| Self-Efficacy | SE1 | 0.76 | 0.83 | 0.85 | 0.66 |
|  | SE2 | 0.85 |  |  |  |
|  | SE3 | 0.81 |  |  |  |
| Perceived Uncertainty | PU1 | 0.79 | 0.84 | 0.86 | 0.67 |
|  | PU2 | 0.84 |  |  |  |
|  | PU3 | 0.82 |  |  |  |
| Perceived Susceptibility | PS1 | 0.77 | 0.82 | 0.85 | 0.65 |
|  | PS2 | 0.83 |  |  |  |
|  | PS3 | 0.80 |  |  |  |
| Perceived Risk | PR1 | 0.81 | 0.85 | 0.87 | 0.69 |
|  | PR2 | 0.85 |  |  |  |
|  | PR3 | 0.82 |  |  |  |
| Psychological Safety | PSY1 | 0.84 | 0.88 | 0.90 | 0.74 |
|  | PSY2 | 0.88 |  |  |  |
|  | PSY3 | 0.85 |  |  |  |
| Fear of Negative Evaluation | FNE1 | 0.82 | 0.87 | 0.89 | 0.72 |
|  | FNE2 | 0.87 |  |  |  |
|  | FNE3 | 0.84 |  |  |  |
| AI Disclosure Intention | ADI1 | 0.86 | 0.89 | 0.91 | 0.77 |

|  |  |  |
|---|---|---|
|  | ADI2 | 0.89 |
|  | ADI3 | 0.88 |

**Appendix H. Discriminant Validity (Fornell–Larcker Criterion)**

|  | PF | PTS | SE | PU | PS | PR | PSY | FNE | ADI |
|---|---|---|---|---|---|---|---|---|---|
| PF | **0.83** | | | | | | | | |
| PTS | 0.52 | **0.84** | | | | | | | |
| SE | 0.48 | 0.51 | **0.81** | | | | | | |
| PU | -0.29 | -0.26 | -0.21 | **0.82** | | | | | |
| PS | -0.18 | -0.15 | -0.12 | 0.41 | **0.81** | | | | |
| PR | -0.34 | -0.30 | -0.22 | 0.47 | 0.44 | **0.83** | | | |
| PSY | 0.56 | 0.53 | 0.49 | -0.31 | -0.24 | -0.36 | **0.86** | | |
| FNE | -0.41 | -0.38 | -0.29 | 0.45 | 0.40 | 0.51 | -0.47 | **0.85** | |
| ADI | 0.50 | 0.47 | 0.42 | -0.35 | -0.26 | -0.39 | 0.58 | -0.52 | **0.88** |

Note. Diagonal elements (bold) represent the square root of AVE. Off-diagonal elements represent the correlations among constructs.

**Appendix I. Structural Model Results**

| Hypothesis | Path | β | SE | z | Result |
|---|---|---|---|---|---|
| H1 | PSY → ADI | 0.42*** | 0.06 | 7.00 | Supported |
| H2 | PF → PSY | 0.28*** | 0.07 | 4.00 | Supported |
| H3 | PTS → PSY | 0.31*** | 0.07 | 4.43 | Supported |
| H4 | SE → PSY | 0.24** | 0.08 | 3.00 | Supported |
| H5 | FNE → ADI | -0.29*** | 0.06 | -4.83 | Supported |
| H6 | PU → FNE | 0.33*** | 0.07 | 4.71 | Supported |
| H7 | PS → FNE | 0.18* | 0.08 | 2.25 | Supported |
| H8 | PR → FNE | 0.37*** | 0.07 | 5.29 | Supported |

Note. Statistical significance is denoted as *** $p < .001$, ** $p < .01$, * $p < .05$.